\documentclass[a4paper]{jpconf}
\usepackage{graphicx,amsmath, amssymb,amsbsy,graphicx,amsthm,color,txfonts}

\input{Qcircuit}
\providecommand{\ket}[1]{\vert#1\rangle}
\providecommand{\bra}[1]{\langle#1\vert}
\newcommand{\kebra}[1]{| #1\rangle\!\langle #1 |}

\begin{document}
\title{The geometric approach to quantum correlations: Computability versus reliability}

\author{$^1$Tommaso Tufarelli,  $^{2}$Tom MacLean,  $^{2}$Davide Girolami, $^{3}$Ruggero Vasile, and $^{2}$Gerardo Adesso}

\address{$^1$QOLS, Blackett Laboratory, Imperial College London, London SW7 2BW, United Kingdom}
\address{$^2$School of Mathematical Sciences, The University of Nottingham, University Park, Nottingham NG7 2RD, United Kingdom}
\address{$^3$Instituto de Fisica Interdisciplinar y Sistemas Complejos (IFISC), Universitar de las Illes Balears, E-07010 Palma de Mallorca, Spain}

\ead{tommaso.tufarelli@gmail.com}

\begin{abstract}
We propose a modified metric based on the Hilbert-Schmidt norm and adopt it to define a rescaled version of the geometric measure of quantum discord. Such a measure is found not to suffer from the pathological dependence on state purity. Although the employed metric is still noncontractive under quantum operations, we show that the resulting indicator of quantum correlations is in agreement with other {\it bona fide} discord measures in a number of physical examples. We present a critical assessment of the requirements of reliability versus computability when approaching the task of quantifying, or measuring, general quantum correlations in a bipartite state.
\end{abstract}
\section{Introduction}

In recent years, a remarkable effort has been dedicated to study the application of quantum mechanical laws to increasingly complex systems. Entanglement undoubtedly takes the central stage in the quest for defining and realizing quantum enhanced tasks, and is generally believed to be {\it the} prominent resource for quantum information processing \cite{ent}. Yet in realistic implementations of quantum technologies involving mixed states, some types of quantum correlations (QCs) can be revealed beyond and even without entanglement, such as those associated to the notion of  {\it quantum discord} \cite{OZ,HV,horo,perinotti}. Such correlations are present in any state $\rho$ of a composite system which cannot be described by classical probability theory, i.e., which is not an embedding of a classical probability distribution. States with nonzero discord are ubiquitous \cite{acin}:  a timely question becomes thus that of finding meaningful approaches to {\it quantify} the amount of discord-type QCs in the quantum states of a system under investigation. 

The recent literature on the field contains already a vast zoology of proposals for measures discord-type QCs \cite{revmodi}.  One can group them in two main classes: entropic quantities, such as the original quantum discord itself \cite{OZ,HV},  and geometric ones, such as the {\it geometric discord} introduced in \cite{dakic}. An entropic measure is usually introduced to provide a thermodynamical interpretation of QCs \cite{szil,deficit}, or to link them to the performance of some quantum protocol \cite{datta,eastin,genio,dattamerging,streltsov,np2}. The geometric measures of QC are instead constructed by fixing a metric in the Hilbert space, and then using it to evaluate the distance between the state under examination and the set of classical (zero discord) states. In the literature, the term {\it geometric discord} is commonly associated to the use of the Hilbert-Schmidt metric \cite{dakic,luofu}, and such QC quantifier has been appreciated for its computability and experimental accessibility. In addition, the geometric discord has been linked quantitatively to the performance of the remote state preparation protocol for two-qubit states  \cite{np1}, and of the same protocol in the context of hybrid resources (qubit {\it vs} harmonic oscillator), when only unitary corrections are allowed on the remotely prepared state \cite{tufo}.
Yet, as timely commented in \cite{piani}, the most commonly used geometric measure of discord suffers two relevant pathologies which endanger its role as a reliable QC quantifier. The first one appears when we allow the unmeasured party to undergo a non-unitary evolution (described by a completely positive local operation). It is observed that geometric discord can increase under this kind of operation \cite{cinesi}, in contrast to what happens when the entropic measure is picked \cite{streltsov}. Technically, such a property is directly related to the non-contractivity of the Hilbert-Schmidt norm \cite{ozawa}. This implies that, strictly speaking, the geometric discord should only be interpreted as a {\it lower bound} to a well-behaved measure of QCs \cite{piani}. The second `bug' arises when one deals with high dimensional systems \cite{tufo,piani}. It is found that highly mixed states containing non-zero and even near-maximal QCs as measured by entropic discord may present negligible geometric quantum discord \cite{tufo,gauss,ginanuovo,martino}. Again, this is due to the properties of the Hilbert-Schmidt distance, which is highly sensitive to state purity. The implication is that geometric discord in high dimensions may fail to be useful even as a lower bound for QC. As a consequence, in such a case, the information-theoretical interpretation of geometric discord, and more generally the relationship between the efficiency of selected quantum protocols and QC measures, may be overrated when not misunderstood.  It is worth to mention that both problems may be fixed by choosing a more suitable and mathematically well behaved metric, such as the trace distance \cite{assistTaka,sarandy} or the Bures distance \cite{ben,faber,orszag}. Unfortunately, this means in most cases that explicit computability has to be given up \cite{revmodi}.

In this paper, we propose an improved recipe to use the Hilbert-Schmidt norm to define a distance in state space, and hence a geometric measure of QCs. Our main aim is to prevent the QC measure from being biased by the global purity of the state, hence answering the main concern reported in \cite{piani}, while still exploiting the low computational demands of the Hilbert-Schmidt norm. The method adopted is the following: first, we define a Hilbert space metric which is based on the Hilbert-Schmidt norm, but is not bounded by the purity of the input states; then, we derive a computable QC geometric quantifier based on such distance, and study its behaviour in a number of explicit examples where the original geometric discord is known to be meaningless. In all the considered case studies, we observe that the newly introduced quantity behaves similarly to the entropic discord, and it could thus serve as a meaningful indicator of discord-like QCs. However, the new measure still inherits the non-contractive behaviour under quantum channels. We thus argue that, as a means to provide analytical insight onto the behaviour of discord-like correlations, the new measure is always preferable to the original geometric discord, yet  more investigations are required to evaluate the extent to which the former is reliable beyond the presented examples.

The paper is organized as follows. In section 2 we review the definitions and main properties of the two main  QC measures, the entropic and geometric discord, discussing the current related issues. In section 3 and subsections, we introduce a  distance in the space of quantum states, from which we derive a computable QC geometric measure, called `rescaled geometric discord'. In Section 4 we study the reliability of the newly introduced measure in a number of relevant case studies, showing that it is able to overcome some of the pathologies associated to the original geometric discord. A comparison between the rescaled geometric discord and entanglement measured by the negativity is discussed in Section 5. Our conclusions are presented in Section 6.  In Appendix A we discuss the evaluation of the new measure in $2\times d$ and $d_A\times d_B$ systems, while the extension of our measure to continuous variable Gaussian states is discussed in Appendix B.

\section{Quantum discord and geometric quantum discord}
We denote by $\rho $ a generic bipartite quantum state with support in a Hilbert space with tensor product structure $\mathcal{H}_{AB}=\mathcal{H}_{A}\otimes\mathcal{H}_{B}$. The entropic quantum discord \cite{OZ,HV} measures the amount of correlations, between the subsystems, that is lost by making a local measurement, say on the system $A$. The explicit formula reads
\begin{equation}\label{ZurDis}
\mathcal{D}^{A}(\rho)=\mathcal{S}(\rho_A)-\mathcal{S}(\rho )+\min_{\{\Pi_j\}}\sum
p_j\mathcal{S}\left(\rho^{\Pi_j}_{B|A}\right),
\end{equation}
where $\mathcal{S}(\cdot)$  indicates the Von Neumann entropy, $\rho_A={\rm Tr}_B\{\rho\}$, $\{\Pi_j\}$ denotes a complete set of projectors in $\mathcal{H}_A$, with $\Pi_j \Pi_k=\Pi_j\delta_{jk}, \sum_j \Pi_j=\mathbb{I}_A$, $p_j={\rm Tr}\{\rho \Pi_j\}$, $\rho^{\Pi_j}_{B|A}={\rm Tr}_{A}\{\Pi_j \rho \Pi_j\}/p_j$. The minimization of the conditional entropy is typically performed over all possible rank-1 complete projectors sets, while in some cases it may be extended to generalized measurements (POVM). It is found that discord vanishes for the so called classical-quantum states, which take the form
\begin{equation}\label{ClaQuan}
\chi=\sum p_i|i\rangle_A\langle i|\otimes\rho_B^i,
\end{equation}
where $\sum_i p_i =1$ and $\{|i\rangle\}$ is an orthonormal vector set. Quantum discord does not increase under local operations---i.e., completely positive and trace preserving (CPTP) maps---on the unmeasured party $B$, that is
\begin{equation}\label{Contrac}
\mathcal{D}^A([\mathbb{I}_A\otimes\Lambda_B]\rho )\leq\mathcal{D}^{A}(\rho ).
\end{equation}
Conversely, CPTP maps on subsystem $A$ can increase and even create quantum discord when applied to classically correlated states \cite{strocal,cicgiova}. From now on we shall omit the label $A$, assuming that measurements are always performed on subsystem $A$. The monotonicity in Eq.~(\ref{Contrac}) has been recognized as a requirement for all {\it bona fide} measures of QCs beyond entanglement \cite{revmodi,streltsov,beyond}. 

Due to the hard optimization problem inherent to the definition of quantum discord, no general analytic expression has been found so far, even for two-qubit states \cite{anal}; an exception is represented by two-mode continuous variable Gaussian states, where entropic discord (constrained to Gaussian measurements) \cite{GioPa} is computable \cite{AdeDat}. This is one of the reasons that led to the introduction of geometric QC measures whose evaluation, in some cases, poses lower computational demands as compared to the entropic discord. The latter measures have been employed to study the dynamics of QCs in open quantum systems and in general in problems where an intensive data analysis is required \cite{revmodi}.

For a bipartite state $\rho$, the geometric discord  with measurements on $A$ \cite{dakic} was originally defined as the distance between the state and the set of classical-quantum states of the form of Eq.~\eqref{ClaQuan}. Alternatively, it can be defined as the minimum (squared) distance between the state and the set of `post-measurement' states obtained after a local projective measurement:
\begin{equation}\label{GeoDisc2}
\mathcal{D}_G(\rho )=\alpha_A\min_{\Pi} \| \rho -\Pi[\rho ]\|^2,
\end{equation}
where $\alpha_A$ is a normalization constant depending on the dimension of $\mathcal{H}_A$, $\Pi[\rho ]= \sum_i\Pi_i\rho \Pi_i$ is the post-measurement state, and $\{\Pi_i\}$ is a complete set of rank-1 projectors on $A$, as above. We shall adopt the convention
\begin{equation}\label{alfano}
\alpha_A=\frac{d_A}{d_A-1}\,,
\end{equation} where $d_A={\rm dim}\{\mathcal H_A\}$.
%, so that $\mathcal{D}_G$ reaches unity on pure, maximally entangled states.
When the norm (and hence the distance) used is induced by the Hilbert-Schmidt scalar product, i.e. $\|A\|=\sqrt{{\rm Tr}\{A^{\dag}A\}}$, the two definitions are equivalent \cite{luofu}, while in general the two geometric approaches may yield different results. Just as the entropic discord, the geometric measure vanishes for classical-quantum states and can increase when a CPTP map is applied to the measured subsystem. Moreover, due to the choice of the Hilbert-Schmidt metric, its evaluation poses fewer difficulties as compared to the entropic discord: closed analytical expressions have been provided for general states of $2\times d$ systems and two-mode Gaussian states \cite{dakic,luofu,rau,gharibian,pirla,gauss}.
On the other hand, the choice of the Hilbert-Schmidt metric is at the heart of the `pathologies' pointed out in  \cite{piani}. One can note that quantum states with different purities have a different Hilbert-Schmidt norm: $\|\rho\|=\sqrt{{\rm Tr}\{\rho^2\}}$. Then, the induced metric is bounded by the input purities as
\begin{equation}
	\|\rho_1-\rho_2\|\leq\|\rho_1\|+\|\rho_2\|=\sqrt{{\rm Tr}\{\rho_1^2\}}+\sqrt{{\rm Tr}\{\rho_2^2\}}.
\end{equation}
This implies that the employed metric does not yield reliable information about the distinguishability of mixed states. This problem becomes particularly prominent in a high-dimensional Hilbert space: one can easily construct a pair of highly mixed states with mutually orthogonal supports, leading to the paradoxical situation in which perfectly distinguishable states appear to be very close to each other in Hilbert-Schmidt norm. In the context of QCs, this brings about two major issues: first, it is possible to construct highly discordant (even entangled) states that possess vanishing geometric discord \cite{tufo}; second, the geometric discord becomes ill-defined when ancillary systems are considered, even if the latter are completely uncorrelated with the main system \cite{piani}. Additionally, as already pointed out, the geometric discord is not monotonically decreasing under CPTP maps on the subsystem $B$: one can find states and channels such that ${\cal D}_G(\rho )\leq {\cal D}_G([\mathbb{I}\otimes\Lambda]\rho )$ \cite{cinesi}.

\section{Proposed deformation of the Hilbert-Schmidt distance and rescaled discord}
In this section, we propose an alternative metric in the state space for the evaluation of the distance between two density matrices. Subsequently, we shall apply such metric to the computation of geometric discord. Our aim is to preserve the low computational demands of the Hilbert-Schmidt metric, while at the same time avoiding its sensitivity to the mixedness of the input states. To treat states of different purities on the same footing, we  propose to normalize each state by its Hilbert-Schmidt norm. Hence, given two density matrices $\rho_1,\rho_2$, we define their distance as
\begin{equation}\label{TufoDist}
d_T(\rho_1,\rho_2)\equiv\biggl\|\frac{\rho_1}{\|\rho_1\|}-\frac{\rho_2}{\|\rho_2\|}\biggl\|,
\end{equation}
where $\|\cdot\|$ indicates, once again, the Hilbert-Schmidt norm. It is trivial to prove that the expression \eqref{TufoDist} is positive, symmetric and satisfies the triangular inequality. Moreover, the additional constraint of $\rho_1,\rho_2$ being density matrices implies that $d_T(\rho_1,\rho_2)=0\Leftrightarrow\rho_1=\rho_2$. This means that $d_T$ is a well defined metric in the space of quantum states. We may then use it to define the ``{\it rescaled} geometric discord'' ({\it rescaled discord} for brevity) $\mathcal{D}_T(\rho )$, by modifying Eq.~\eqref{GeoDisc2} in the following way
\begin{equation}\label{TufGeoDisc}
\mathcal{D}_T(\rho )=\beta_A\min_{\Pi}d_T(\rho ,\Pi[\rho ])^2,
\end{equation}
with $\beta_A$ a normalization constant which is again dependent on the
dimension of $\mathcal{H}_A$. If we restrict the minimization to projective measurements, it is possible to connect the two geometric measures in an elegant way. (Generalized measurements will only be considered when dealing with Gaussian states, see \ref{gaussian}). Eq.~\eqref{GeoDisc2} involves the
minimization of the quantity
\begin{equation}\label{HS}
{\cal Q}_{\rm HS}=\|\rho -\Pi[\rho]\|^2={\rm Tr}\{\rho^2\}-{\rm Tr}\{\rho\Pi[\rho]\},
\end{equation}
where the second equality requires that
${\rm Tr}\{\rho\Pi[\rho]\}={\rm Tr}\{\Pi[\rho]^2\}$, valid only in the
projective case. On the other hand, in Eq.~\eqref{TufGeoDisc} we have to minimize the quantity
\begin{equation}\label{Equiv1}
{\cal Q}_T=d_T(\rho,\Pi[\rho])^2=2-2\sqrt{\frac{{\rm Tr}\{\rho\Pi[\rho]\}}{{\rm Tr}\{\rho^2\}}}=
2-2\sqrt{1-\frac{{\cal Q}_{\rm HS}}{{\rm Tr}\{\rho^2\}}}.
\end{equation}
The last expression states that, keeping fixed the initial state $\rho$, the quantity ${\cal Q}_T$ is a monotonically increasing function of ${\cal Q}_{\rm HS}$. Hence, the projective measurements
minimizing the geometric and the rescaled discord are the
same, independently on the dimensionality of the system, and from Eq.~\eqref{Equiv1} it follows that
\begin{equation}\label{Equiv2}
\min_{\Pi}{\cal Q}_T=2-2\sqrt{1-\frac{\mathcal{D}_G(\rho)}{\alpha_A
{\rm Tr}\{\rho^2\}}}.
\end{equation}
Incidentally, this also implies that minimizing Eq.~\eqref{TufGeoDisc} is equivalent to minimizing the distance between the state $\rho$ and the set of classical-quantum states of the form \eqref{ClaQuan}, as it was the case for the original geometric discord \cite{luofu}.  We can then employ the complete expression \eqref{TufGeoDisc}, conveniently normalizing it such that the geometric and rescaled discord are equal
for pure, maximally entangled states. In that case we have
\begin{equation}\label{Costnorm}
\beta_A=\frac{\mathcal{D}^{\max}_G}{2-2\sqrt{1-\mathcal{D}^{\max}_G/\alpha_A}}
\end{equation}
where $\mathcal{D}^{\max}_G$ is the value of the geometric discord for
a maximally entangled state, equal to $1$ if Eq.~({\ref{alfano}) is adopted. It follows that
\begin{equation}\label{NormDisc2}
\mathcal{D}_T(\rho)=\beta_A\left[2-2\sqrt{1-\frac{\mathcal{D}_G(\rho)}{\alpha_A{\rm Tr}\{\rho^2\}}}\,\right].
\end{equation}
The above expression shows that the rescaled discord is obtained effectively by renormalizing the original geometric discord by the purity of the input state. For pure states,   ${\rm Tr}\{\rho^2\}=1$ and hence the rescaled discord becomes simply a function of the geometric discord, and reduces in particular to an entanglement monotone like the latter.

One can notice that it was enough to adjust the geometric discord by just dividing it by the state purity, in order to overcome the main pathology pointed out in \cite{piani}, namely the fact that ${\cal D}_G$ can be changed by reversible operations on the unmeasured party (like appending a mixed ancilla). The so-adjusted geometric discord, definable as
\begin{equation}\label{NormDisc1}
\widetilde{\mathcal{D}}_G(\rho)=\frac{\mathcal{D}_G(\rho)}{{\rm Tr}\{\rho^2\}},
\end{equation}
is a simple monotonic function of the rescaled discord ${\cal D}_T$ of Eq.~(\ref{NormDisc2}). We remark that rescaling quantities based on the Hilbert-Schmidt metric by the state purity has been already considered in other information-theoretical settings \cite{winter,genoni}, where such an operation has been regarded as correcting for the `effective dimension' of the involved states \cite{winter}.

However, as a matter of fact the rescaled geometric measure inherits from $D_G$ the non-contractivity problem under CPTP evolutions on the unmeasured system: in particular, the counterexamples in \cite{cinesi} affect ${\cal D}_T$ and ${\widetilde{\cal D}}_G$ as well. This means that the newly found quantity is again to be regarded as an indicator rather than as a full-fledged measure of QCs. One might define such a {\it bona fide} measure by introducing a maximization for  ${\cal D}_T$ over all CPTP operations on the unmeasured subsystem $B$, as suggested in  \cite{piani}, although this would render the corresponding measure practically uncomputable. Nevertheless, the case studies provided in ~\ref{casestud} suggest that our lower bound ${\cal D}_T$ may be a meaningful estimator of the QC content of bipartite states, regardless of their purity and of the dimensionality of the systems involved. We reiterate that any analytical insight based on ${\cal D}_T$ and ${\widetilde{\cal D}}_G$ should be backed up, even if just numerically, by parallel studies based on fully well-behaved measures, such as the entropic discord or geometric measures based on contractive distances.

\section{Case studies}\label{casestud}
In this section we investigate the reliability of the rescaled discord in estimating the QC content of bipartite states. To do so, we shall consider three case studies which are well established as being exemplary test-beds for any meaningful QC quantifier. We shall see that, in each of the selected cases, the rescaled discord ${\cal D}_T$ is a good indicator of the QC content of the considered states, reproducing their behaviour as measured by the entropic discord. In contrast, the unrescaled geometric discord $D_G$ progressively loses reliability as the mixedness of the considered states is increased.
\subsection{DQC1 model}
In our first case study we compare entropic and geometric QC quantifiers in a non-universal model of quantum computation. The DQC1 (deterministic quantum computation with one clean bit) \cite{laf1} is a protocol which estimates the trace of a unitary matrix, say $U$. This is a paradigmatic case for the study of QCs: fixing the desired accuracy of the estimation, discordant states may provide exponential speed up with respect to classical ones, whilst entanglement is negligible over all the computation \cite{laf1}. In this sense, quantum discord was proposed as the possible explanation behind the power of DQC1 \cite{datta}.

The model is described as follows. The resources are an ancillary qubit $\rho_A=\frac 12(\mathbb{I}_2+ \mu \sigma_z)$ and a register of $n$ qubits in a maximally mixed state, $\rho_B=\frac 1{2^n}\mathbb{I}_n$.  They are initially in a product state $\rho_{\text{in}} =\rho_A\otimes \rho_B$. Firstly, a Hadamard gate is applied to the ancilla, followed by a global control-$U$ operation. Then, spin measurements are made on the ancilla, and from their outcomes one can retrieve information on the trace of the matrix $U$. The scheme can be described by the following circuit
\begin{eqnarray}\label{dqc1}
 \Qcircuit @C=1.4em @R=1.2em
{
\lstick{\frac 12(\mathbb{I}_2+ \mu \sigma_z)} &  \gate{H} &  \ctrl{1}  &\meter \\
\lstick{\mathbb{I}_n/2^n} &\qw & \gate{U} &\qw\\
}
\end{eqnarray}
The output state, before the measurement, may be expressed in the ancillary qubit basis as
 \begin{eqnarray}
\rho_ {\text{out}}&=&\frac1{2^{n+1}}\left(
\begin{array}{c|c}
 \mathbb{I}_{n} & \mu U^{\dagger}  \\ \hline
 \mu U &   \mathbb{I}_{n} \\
\end{array}
\right),
  \end{eqnarray}
   while the final (reduced) state of the ancilla reads
     \begin{eqnarray}
  \rho_A^ {\text{out}}&=&\frac1{2}\left(
\begin{array}{cc}
 1 & \mu \text{Tr}[U^{\dagger}]  \\
 \mu \text{Tr}[U] &   1 \\
\end{array}
\right).
  \end{eqnarray}
It is immediate to see that: $\langle\sigma_1\rangle_{\rho_A^{out}}=\mu\ \text{Re}\left[\text{Tr}[U]\right], \langle\sigma_2\rangle_{\rho_A^{out}}=\mu\ \text{Im}\left[\text{Tr}[U]\right]$. This shows that, as anticipated, the efficiency of the protocol (keeping fixed the desired accuracy on the estimation of ${\rm Tr}[U]$) is solely dependent on the ancilla initial polarization $\mu$, and not on the dimensionality of the unitary $U$.

 The QCs between ancilla and register in the output state have been studied by using both entropic and geometric measures at both theoretical and experimental level \cite{datta,white,laf2,ginanuovo,revmodi}. It was found that entropic discord is independent of the dimension of the system, as a reliable figure of merit of the protocol should be, while geometric discord monotonically decreases by increasing the number of qubits in the register. As extensively discussed, this is due to the inadequacy of the Hilbert-Schmidt metric in distinguishing mixed states. Now, we can study the behaviour of a properly normalized geometric measure as given by Eq.~\eqref{NormDisc1} or \eqref{NormDisc2}, and compare it to the entropic and geometric discord. For the entropic discord, one has the approximate expression \cite{datta} $${\cal D}(\rho_{\text{out}})\sim 2+h\left\{\frac{1-\mu}{2}\right\}+h\left\{\frac{1+\mu}{2}\right\}-\log_2\left\{1-\sqrt{1-\mu^2}\right\}- \left(1-\sqrt{1-\mu^2}\right)\log_2\{e\},$$ where $h\{x\}=x \log_2\{x\}$. Easy calculations return the values of the geometric discord and consequently of its normalized counterparts \cite{pirla}. We study the evolution of the four quantities %and a further probe measure ${\cal D}_G/{\rm Tr}\{\rho_{\text{out}}\}$
by varying the polarization of the ancilla in Fig.~\ref{dqcfig1} (Left). The geometric discord reveals its dependence on the purity of the global state, which makes its evolution depressed with respect to the evolution of the other quantities. Then, in Fig.~\ref{dqcfig1} (Right), we fix the ancilla polarization and modulate the number of qubits in the register. It is found that our proper rescaling allows to associate even a geometric QC measure, i.e., $\mathcal{D}_T$ or $\widetilde{\mathcal{D}}_G$, with the performance of the protocol.
\begin{figure}[tb]
	\begin{center}
\includegraphics[width=0.9\textwidth]{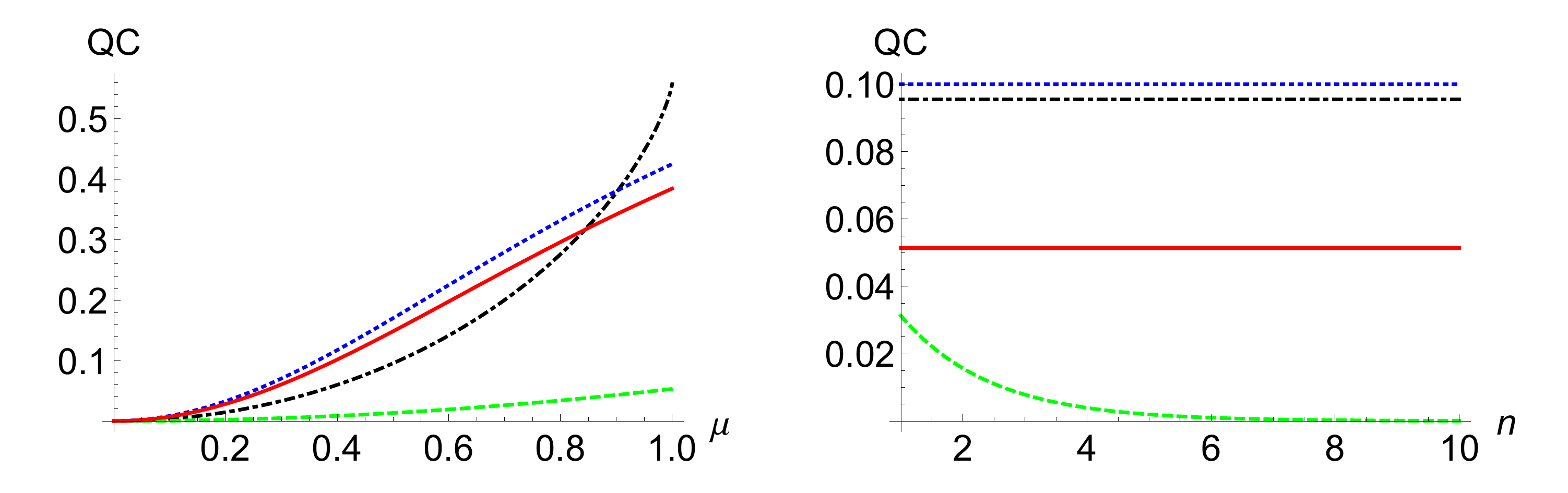}
\caption{\label{dqcfig1}(Left) DQC1 model: QCs between ancilla of polarization $\mu$
and the three-qubit register in the output state. We consider the implementation of \cite{laf2}, where the register consists of  a three-qubit state and the unitary is given by $U ={\rm diag} (a, a, b, 1, a, b, 1, 1)$, with $a=-({\rm e}^{-i3\pi/5})^4$, $b=({\rm e}^{i3\pi/5})^8$. Green dashed line:
geometric discord ${\cal D}_G$; Black dot-dashed line: entropic discord ${\cal D}$; Red line: rescaled discord $\mathcal{D}_T$;  Blue dotted line: $\widetilde{\mathcal{D}}_G={\cal D}_G/ {\rm Tr}\{\rho_{\text{out}}^2\}$. %The increasing global purity keeps the geometric discord well below the other measures.
(Right) DQC1 model: here we fix the ancilla polarization being $\mu=0.5$ and
vary the dimension of the register, i.e., the number of qubits $n$. We consider unitaries $U$ such that ${\rm Tr}[U]={\rm Tr}[U^2]=0$, with uniformly distributed eigenvalues, which allow for the analytical evaluation of entropic and geometric discord \cite{datta}. Green dashed line: geometric discord ${\cal D}_G$; Black dot-dashed line: entropic discord ${\cal D}$; Red line: rescaled discord $\mathcal{D}_T$;  Blue dotted line: $\widetilde{\mathcal{D}}_G={\cal D}_G/ {\rm Tr}\{\rho_{\text{out}}^2\}$. It is immediate to appreciate the peculiar decay of geometric discord for high dimensions, while the entropic and rescaled discord are constant.}
\end{center}
\end{figure}
\subsection{Qubit-oscillator states}
We consider here a family of correlated states of one qubit $A$ and one continuous variable harmonic oscillator $B$ of the form
\begin{equation}
	\rho\!=\!p\kebra{0}\; D(\beta)\rho_B^0 D^\dagger(\beta)+(1\!-\!p)\kebra{1}\; D^\dagger(\beta)\rho_B^0 D(\beta)+r\ket{0}\bra{1}\; D(\beta)\rho_B^0 D(\beta)+r^*\ket{1}\bra{0}\; D^\dagger(\beta)\rho_B^0 D^\dagger(\beta),\label{qubit-osc}
\end{equation}
where $0\leq p\leq1$, $|r|^2\leq p(1-p)$, $D(\beta)$ is the oscillator's displacement operator and $\rho_B^0$ is a generic oscillator initial state, see \cite{tufo} for details. We shall focus on the limit in which $\rho_B^0$ is highly mixed, that is, ${\rm Tr}\{(\rho_B^0)^2\}\sim0$, and $|\beta| \rightarrow \infty$, which in practice means $|\beta|$ large enough such that the overlap between the two phase-space domains, associated to $\rho_B^0$ displaced by $\beta$, and to $\rho_B^0$ displaced by $-\beta$, becomes negligible. In these limits, analytic results are available for the negativity $\mathcal N$, the geometric discord $\mathcal D_G$, and a meaningful lower bound $\mathcal D^\text{Lo}$ to the entropic discord of these states \cite{tufo}. From the knowledge of $\mathcal D_G$, analytic results for $\widetilde{\mathcal{D}}_G$ and $\mathcal{D}_T$ follow trivially and shall not be presented here for brevity. It is found that the geometric discord (for measurements on the qubit) of the states \eqref{qubit-osc}, in the described parameter regime, is always negligible. On the other hand, the entropic discord and even the negativity are nonzero as far as $r\neq0$, clearly indicating the inadequacy of geometric discord in capturing the nonclassical character of the states of Eq.~\eqref{qubit-osc} \cite{tufo}. In contrast, Fig.~\ref{grigino} shows that the rescaled discord provides nontrivial upper and lower bounds to the quantity $\mathcal D^\text{Lo}$, for all physical values of the parameters $r$ and $p$.
\begin{figure}[tb]
	\begin{center}		
\includegraphics[width=.9\textwidth]{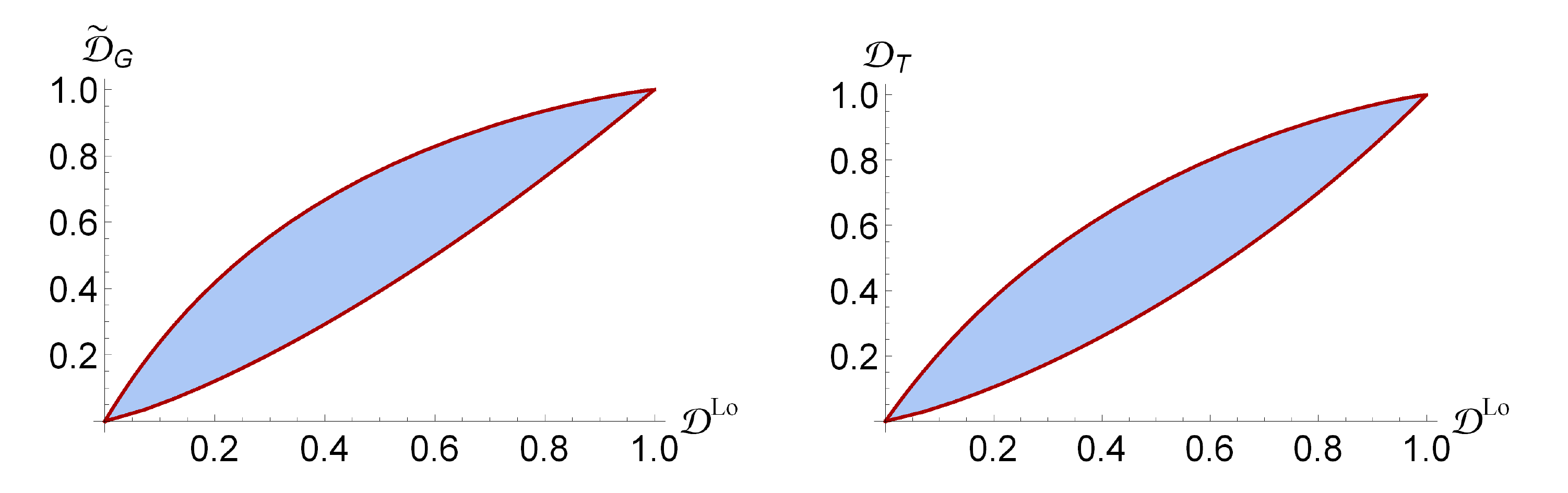}
	\end{center}
	\caption{Measures of QCs in the qubit-oscillator states of Eq.~(\ref{qubit-osc}). (Left) Parametric plot of the quantities $\mathcal D^\text{Lo}$ (horizontal axis) and $\widetilde{\mathcal{D}}_G$ (vertical axis), for all physical values of the parameters $p,r$. (Right) Parametric plot of $\mathcal D^\text{Lo}$ and $\mathcal{D}_T$. The plots show how the rescaled geometric discord provides upper and lower bounds to $\mathcal D^\text{Lo}$, as opposed to the geometric discord $\mathcal D_G$ which is negligible for all values of $p,r$ \cite{tufo}. \label{grigino}}
\end{figure}
\subsection{Werner states of $d\times d$ systems.}
As a final example, we investigate the reliability of the rescaled discord for generic $d\times d$ systems. Due to the complexity of evaluating QCs in high dimensions, we shall limit our investigations to a class of highly symmetric states, which allow for the analytical calculation of both $\mathcal D$ and $\mathcal D_G$. (A lower bound to ${\cal D}_T$ for generic $d_A \times d_B$ states is presented in \ref{appqdqd}).
The {\it Werner states} are defined as
\begin{equation}
	\rho_W(\lambda)=\frac{2(1-\lambda)}{d(d+1)}\Pi^++\frac{2\lambda}{d(d-1)}\Pi^-,
\end{equation}
where $\Pi^\pm\equiv (\mathbb I\pm\mathbb F)/2$ are the projections over the symmetric and antisymmetric subspaces of $\mathbb C^d\otimes\mathbb C^d$ respectively, $\mathbb F$ being the swap operator $\mathbb F(\ket{\phi}\otimes\ket{\psi})=\ket{\psi}\otimes\ket{\phi}$. In terms of the entropic discord, these states enjoy a nonclassical character, regardless of the dimension $d$, for all but one special value of the parameter $\lambda$. The geometric discord, however, tends to underestimate such nonclassicality as the dimension $d$ is increased, due to the corresponding increase in state mixedness \cite{chitambar}. Fig.~\ref{werner} shows how the rescaled discord captures well the qualitative behaviour of the entropic discord of Werner states.

%The {\it isotropic states} are instead obtained by mixing a maximally entangled state with the identity. In particular, we shall use
%\begin{equation}
%	\rho_I(f)=f\ket{\Phi}\bra{\Phi}+\frac{1-f}{d^2-1}(\mathbb I-\ket{\Phi}\bra{\Phi}),
%\end{equation}
%where $\ket{\Phi}=d^{-1/2}\sum_k\ket k\ket k$ and $0\leq f\leq 1$ is the fidelity between $\ket{\Phi}$ and $\rho_I(f)$. The plots in Fig.~(banana) show a comparison of the behaviour of quantities $\mathcal D_G, \mathcal D_Z$, as well as the rescaled discord, as a function of the dimension $d$. We see that, as opposed to the geometric discord, the newly introduced quantity shows a scaling with the system dimension which is compatible with that of the entropic discord.
\begin{figure}[tb]
	\begin{center}
\includegraphics[width=.9\textwidth]{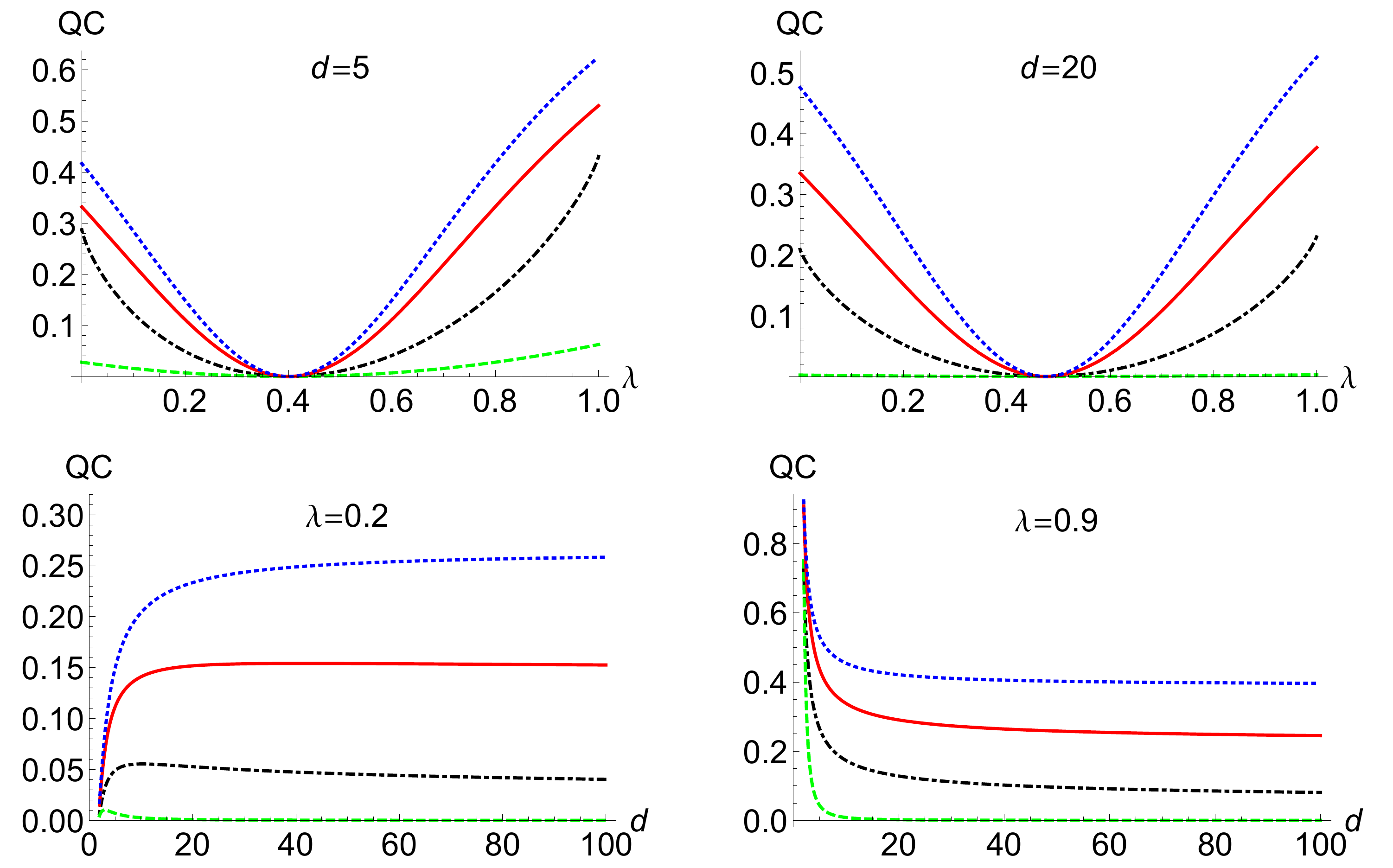}
\end{center}
\caption{Behaviour of QC quantifiers for $d \times d$ Werner states $\rho_W$, as functions of the parameter $\lambda$ and the dimension $d$. Green dashed line: geometric discord ${\cal D}_G$; Black dot-dashed line: entropic discord ${\cal D}$ (normalized by $\log_2 d$); Red line: rescaled discord $\mathcal{D}_T$;  Blue dotted line: $\widetilde{\mathcal{D}}_G={\cal D}_G/ {\rm Tr}\{\rho^2\}$. The top two plots are obtained by fixing the dimension $d$ and varying $\lambda$: in this case the three curves $\mathcal D,\mathcal D_T, \widetilde{\mathcal D}_G$ have a qualitatively similar behaviour, while $\mathcal D_G$ shows little deviation from zero, especially in high dimensions. In the bottom plots $\lambda$ is kept fixed and the dimension $d$ is varied. The geometric discord becomes rapidly negligible with increasing dimension, regardless of $\lambda$, while the other three quantities stabilize to a constant nonzero value which is different for each quantifier and depends on $\lambda$. \label{werner}}
\end{figure}

\section{Relationship between rescaled discord and entanglement}
In this section, we investigate the relationship between the rescaled discord and entanglement, in the form of negativity \cite{ent}, in the spirit of \cite{interplay}.
It is often said that QCs go beyond entanglement. This statement can be made quantitative in several cases. In general, for every entanglement monotone $E$, Ref.~\cite{beyond} proves that there exists a valid measure ${\cal D}_E$ of QCs which reduces to $E$ for pure bipartite states, and satisfies the hierarchy ${\cal D}_E \geq E$ for mixed bipartite states. Unfortunately, these measures are often hardly computable, as it is the case for those quantifiers based on the relative entropy distance \cite{modi}. In Ref.~\cite{interplay}, favoring as in this paper measures enjoying computability in relevant cases, it was shown that a simple hierarchy between QCs and entanglement can be proven for general states of two-qubit systems, involving the geometric discord ${\cal D}_G$ \cite{dakic} and the squared negativity ${\cal N}^2$ \cite{ent}, respectively. Namely, ${\cal D}_G(\rho) \geq {\cal N}^2(\rho)$ for $\rho \in \mathbb C^2 \otimes \mathbb C^2$, with equality on pure states. Initially conjectured for higher dimensional states as well, such a hierarchy has been disproven already in $2 \times 3$ systems \cite{rana}, while suitable modifications of it hold \cite{vianna}. We recall that the negativity, normalized to $1$, can be defined as
\begin{equation}\label{nega}
{\cal N}(\rho) = \frac{1}{d_A-1} (\|\rho^{t_A}\|_1 -1)\,,
\end{equation}
where $\|M\|_1=\text{Tr}|M| = \sum_i |m_i|$ is the trace norm, and  $\rho^{t_A}$ denotes the partial transpose of the bipartite state $\rho$ with respect to the degrees of freedom of subsystem $A$ \cite{ent}.

\begin{figure}[tb]
	\begin{center}
\includegraphics[width=.329\textwidth]{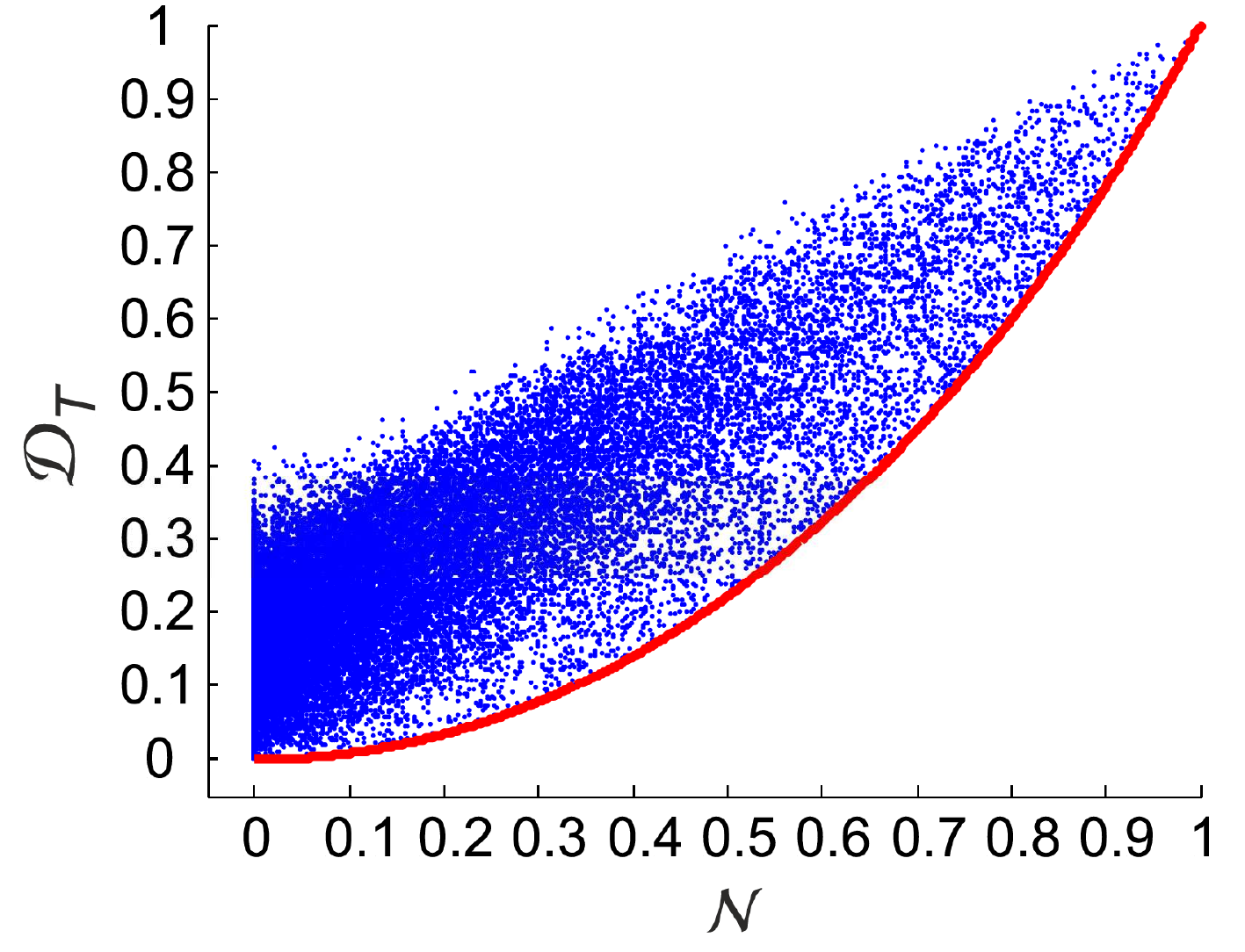}
\includegraphics[width=.329\textwidth]{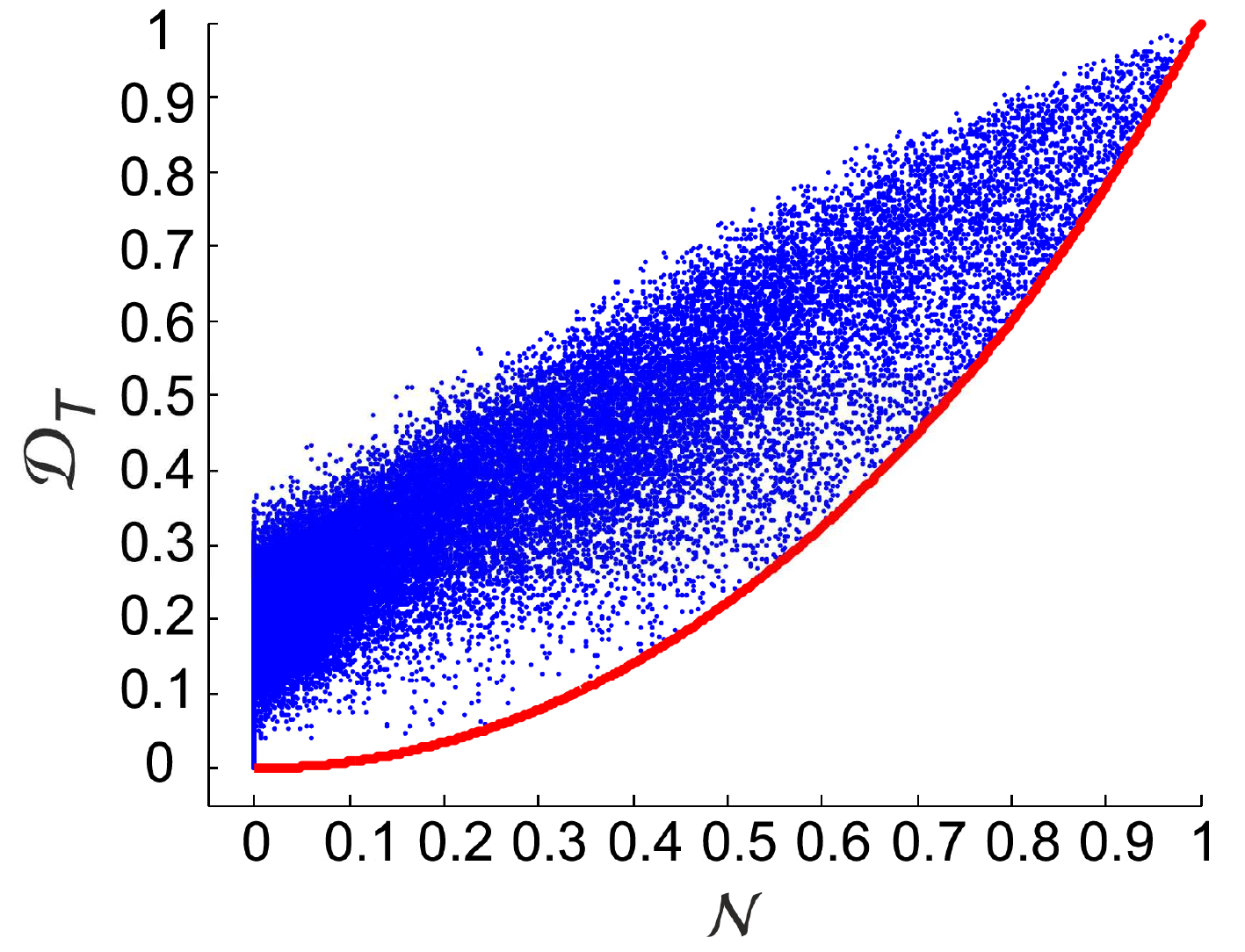}
\includegraphics[width=.329\textwidth]{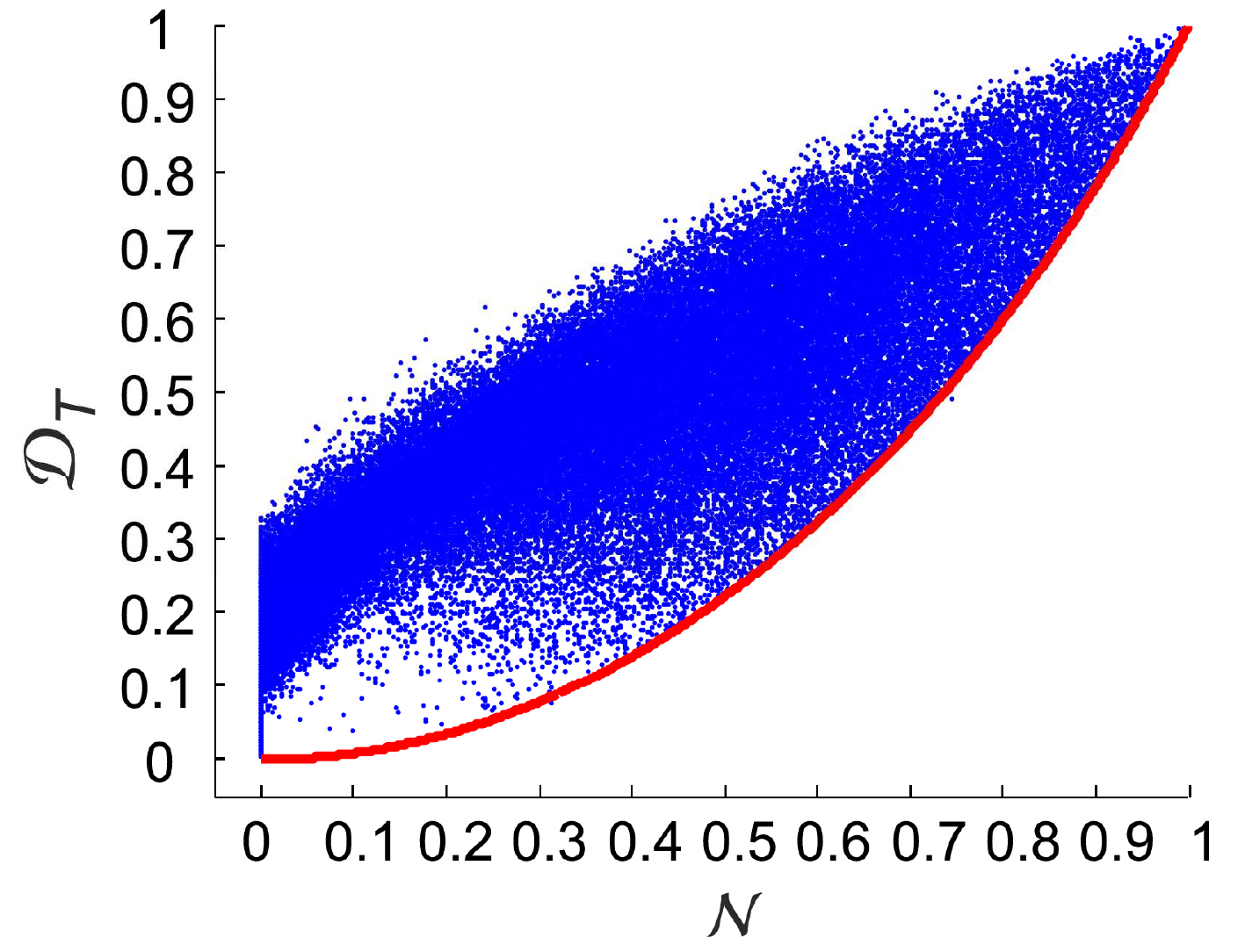}
\end{center}
\caption{Interplay between rescaled discord ${\cal D}_T$ and negativity ${\cal N}$ for states of (from Left to Right) $2\times 2$, $2\times 3$, and $2\times 4$ systems. In each plot, $10^5$ random states are represented, while the red curves accommodate pure states. Some points are visible below the pure-state boundary in the $2 \times 4$ case (Rightmost panel). \label{figtom}}
\end{figure}

It is interesting to study whether the rescaled discord provides an upper bound to the squared negativity for a larger set of states. This is in principle possible, since trivially $\widetilde{\cal D}_G \geq {\cal D}_G$ as the purity is a quantity between $0$ and $1$. We have generated a large sample of random states of $2 \times d$ systems with $d=2,3,4,\ldots$. In Fig.~\ref{figtom}, we plot the complete rescaled ${\cal D}_T$ [Eq.~\eqref{NormDisc2}] versus ${\cal N}$ for such random states. With this choice of quantifiers, the hierarchy would read ${\cal D}_T \geq \left[2-\sqrt{4-2{\cal N}^2}\right]/(2-\sqrt{2})$. The results of \cite{interplay} imply that the rescaled hierarchy is still valid for $d=2$. Interestingly, we have obtained extensive numerical evidence of its validity for $d=3$ as well (whereas ${\cal D}_G \geq {\cal N}^2$ does not generally hold for any $d>2$ \cite{rana}), while we found a few states violating the rescaled hierarchy for $d\geq 4$. These violations correspond to mixed, entangled states whose rescaled discord is smaller than the one of the corresponding pure states with the same degree of negativity. We leave our readers to face the  task of providing analytical corroboration of these findings, perhaps exploiting the methods of \cite{rana,vianna}.

\section{Conclusions}
The study of QCs other than entanglement is still in its infancy \cite{revmodi}. It is important at this stage to highlight which approaches and methods provide reliable results, and which ones lead to an incorrect understanding of the nature and the role of these important nonclassical features of quantum states.
Employing a QC measure based on the Hilbert-Schmidt distance, as the geometric discord, dramatically simplifies calculations---that was indeed the original aim of the authors in \cite{dakic}. On the other hand, the price to pay is a limited reliability and applicative power of the measure. Whenever the purity of the global state is not kept fixed, e.g., the evolution of the system is not unitary, geometric discord cannot be employed as a QC measure \cite{piani}. In this paper, we explored a path to correct for this, by identifying the dependence on state purity as a crucial pathology of the geometric discord. We preferred to maintain computability, and hence proposed a suitable deformation of the Hilbert-Schmidt distance. There are other {\it bona fide} geometric quantifiers, based on contractive distances such as the relative entropy \cite{modi,streltsov,genio}, the trace norm \cite{assistTaka,sarandy}, or the Bures distance \cite{ben,faber,orszag}, which generate reliable measures of QCs, and can properly describe their dynamical evolution, for example when the state is embedded in and interacts with an environment \cite{revmodi}. However, they remain hard to access computationally if not for very special cases. To the best of our knowledge, the only QC quantifier to date that combines {\it both} reliability and computability (the latter in dimension $2\times d$) is the so-called `local quantum uncertainty' introduced very recently in \cite{lqu}.

Our findings suggest that any study where the geometric discord ${\cal D}_G$ is employed without a rescaling by the state purity is likely to return, apart from special circumstances, not entirely trustworthy results even in the restricted setting of two-qubit states. Our rescaled discord ${\cal D}_T$ was here adopted in a number of test cases, returning results in good qualitative and quantitative agreement with what expected based on the scenarios considered and on other known QC measures. However, due to the lack of contractivity of the new metric, we emphasize that further investigations are required to assess, in more quantitative terms, the extent of applicability of such QC indicator.  Very recently, it has been found that all known {\it bona fide} (entropic or distance-based) measures of discord exhibits a peculiar phenomenon under certain dynamical conditions: they stay constant on a class of two-qubit Bell diagonal states when undergoing particular nondissipative decoherence channels up to a critical time. Interestingly, the rescaled discord introduced here side up with the other measures in exhibiting this phenomenon \cite{ben}, unlike the original $D_G$. This extends the reliability of the new quantifier to incorporate paradigmatic features of QCs. 
We believe it could be of interest to investigate further deformations of the Hilbert-Schmidt metric, or closely related metrics, in an attempt to preserve its appealing computability while possibly curing all of its pathologies.

%We believe it could be of interest to construct a hierarchy of quantum, classical and total correlations based on the Tufo metric which defines ${\cal D}_T$, as done in \cite{saruzzo} for the unrescaled Hilbert-Schmidt metric.

\section*{Acknowledgments}
We warmly thank Vittorio Giovannetti and Marco Piani for discussions. This research was supported by the EPSRC under a Vacation Bursary 2012 and the Research Development Fund grants BTG-0612b-31 (2012) and PP-0313-36 (2013).

   \appendix
\section{Evaluation of the rescaled discord}
When only projective measurements are considered, Eqs.~\eqref{NormDisc2} and \eqref{NormDisc1} imply that any analytic result available in the literature for the geometric discord can be exploited straightforwardly to provide the rescaled discord. In the present section, we shall review the available results for the relevant cases of qubit-qudit and qudit-qudit systems.
%, in which general formulas for the rescaled discord can be evaluated explicitly.

%The situation changes, however, when generalized measurements are considered: this is illustrated below for the relevant case of continuous-variable (CV) Gaussian states.
\subsection{Qubit-qudit systems}
We start by analyzing the situation of a bipartite system composed of a qubit and a qudit (where the dimension $d$ of the latter may be finite or infinite), and we shall evaluate the rescaled discord for measurements on the qubit. Here we briefly recall the approach of reference \cite{tufo} which is based on a Bloch expansion of the total state of the composite system as
\begin{equation}\label{StateAB}
\rho =\frac{1}{2}(\mathbf{v}\cdot\boldsymbol\sigma)
\end{equation}
with $\boldsymbol\sigma=(\mathbb{I},\vec{\sigma})=(\mathbb{I},\sigma_x,\sigma_y,\sigma_z)$
being the $2\times 2$ identity operator and the three Pauli
matrices acting on the qubit $A$, while $\mathbf{v}=(v_0,\vec{v})$ is a four dimensional
vector of operators in $\mathcal{H}_B$, defined as
$v_0={\rm Tr}_A\{\rho \}$ and $\vec{v}={\rm Tr}_A\{\vec{\sigma}\rho \}$.
After the introduction of the $3\times 3$ dimensional symmetric
matrix
\begin{equation}\label{Smatrix}
\mathbf{S}={\rm Tr}_{B}\{\vec{v}\,\vec{v}^\intercal\},
\end{equation}
the geometric discord is given by
\begin{equation}\label{GeoDisc2xd}
\mathcal{D}_G(\rho )={\rm Tr}\{\mathbf{S}\}-\lambda_{\max}(\mathbf{S})
\end{equation}
where $\lambda_{\max}(\mathbf{S})$ is the maximum eigenvalue of the
matrix $\mathbf{S}$ associated to the state $\rho $. Upon
substituting Eq.~\eqref{GeoDisc2xd} into \eqref{NormDisc2} we obtain the rescaled discord.

\subsection{Qudit-qudit systems}\label{appqdqd}
%The main issue of determining Zurek's or geometric discord in higher
%dimensional system is the higher number of parameters describing the
%most general projective measurements. Here we study two possible
%ways of tackling this problem. The first one provides a lower bound
%for the rescaled discord, while the second is a
%numerical approach, where we generate random projective measurements
%and use the one that minimize the total quantum correlations. We
%also compare the two situations, actually proving that the first
%method is, in many cases, a strict lower bound for the geometric
%measure.
We show here how to tackle the calculation of the rescaled discord for a generic bipartite system of dimension $d_A\times d_B$. In particular, it is possible to recast the problem as a constrained optimization of a multivariate polynomial, to which we derive an explicit lower bound. From Eq.~\eqref{Equiv1} we deduce that we have to study the maximum of the function ${\rm Tr}\{\rho\Pi[\rho]\}$ over all the possible non-degenerate projective measurements on $\mathcal{H}_A$. We now take a generic bipartite density operator $\rho$ and fix a `computational' basis $\{|1\rangle,|2\rangle...,|d_A\rangle\}$ for the subsystem $A$ where $d_A=\dim\mathcal{H}_A$. A generic density operator may be expressed as
\begin{equation}\label{Qdqd1}
\rho =\sum_{n,m}|n\rangle\langle m|\otimes \rho_{nm}
\end{equation}
where the $\rho_{nm}={\rm Tr}_A[\rho  |m\rangle\langle n|]$ are operators in $\mathcal{H}_B$. A set of complete rank-1 projectors in $\mathcal{H}_A$ may be expressed as $\{P_j=\kebra{\phi_j}\}$, where $\{\phi_j\}_{j=1}^{d_A}$ is an orthonormal basis. Expanding the projectors with respect to the computational basis, we have
\begin{equation}\label{Qdqd2}
P_{j}=\sum_{n,m}|n\rangle\langle
m|(P_j)_{nm}=\sum_{n,m}|n\rangle\langle
m|\phi_{j,n}\phi_{j,m}^*,
\end{equation}
where we have used $|{\phi_j}\rangle=\sum_n\phi_{j,n}\ket n$. Recalling that $\Pi[\rho ]=\sum_jP_j\rho P_j$, we have
\begin{equation}\label{Qdqd3}
{\rm Tr}\{\rho\Pi[\rho]\}=\sum_{j,n,m,p,q}{\rm Tr}_B\{\rho_{nm}\rho_{pq}\}(P_j)^*_{nq}(P_j)_{mp}=\sum_{j,n,m,p,q}{\rm Tr}_B\{\rho_{nm}\rho_{pq}\}\phi_{j,q}\phi_{j,n}^*\phi_{j,m}\phi_{j,p}^*.
\end{equation}
Taking into account the orthonormality of the vectors $\phi_j$, we can recast the problem of finding the optimal projective measurement as the maximization of the quartic form in Eq.~\eqref{Qdqd3}, subject to the $d_A(d_A+1)/2$ constraints
\begin{equation}
		\sum_{n}\phi_{j,n}\phi_{j',n}^*=\delta_{jj'}\qquad j\leq j'=1,2,...,d_A.\label{orthogonals}
\end{equation}
Such a problem may be attacked via the standard  Lagrange multipliers method. Here, we shall be satisfied with finding an explicit lower bound to the rescaled discord in $d_A\times d_B$ systems. To do so, we can reinterpret the matrix elements $(P_j)_{nm}=\phi_{j,n}\phi_{j,m}^*$ as being the components of a column vector of length $d_A^2$. The resulting vectors will be denoted by $\vec{P}_j$, where the arrow helps
in distinguishing the two interpretations. It is easy to check that the orthogonality condition between projectors, $P_jP_k=P_j\delta_{jk}$, implies that in the new representation we have
\begin{equation}\label{Qdqd4}
\vec{P}_j^{\dag}\vec{P}_k=\delta_{jk}.
\end{equation}
With these new definitions we can rewrite Eq.~\eqref{Qdqd3} as
\begin{equation}\label{Qdqd5}
{\rm Tr}\{\rho\Pi[\rho]\}=\sum_j\vec{P}_j^{\dag}\mathbf{A}\vec{P}_j
\end{equation}
where $\mathbf{A}$ is a $d_A^2\times d_A^2$ matrix whose elements are
$\mathbf{A}_{\{n,q\},\{m,p\}}={\rm Tr}_B[\rho_{nm}\rho_{pq}]$. Equation
\eqref{Qdqd5} is a sum of $d_A$ bilinear forms. Each term is bounded
by the largest eigenvalue of the matrix $\mathbf{A}$. However, since the vectors $\vec P_j$ are orthonormal, each eigenvalue may be taken only once, and the maximum of Eq.~\eqref{Qdqd3} is upper bounded by the sum of the $d_A$ largest eigenvalues of the matrix
$\mathbf{A}$, that is,
$\max_{\Pi}{\rm Tr}\{\rho\Pi[\rho]\}\leq\lambda_{1}(\mathbf{A})+\lambda_{2}(\mathbf{A})+\ldots+\lambda_{d_A}(\mathbf{A}).$
The inequality sign is due to the fact that, in general, equality may be reached by vectors $(\vec{P}_j)$ whose $d_A^2$ elements are not in the required form $(P_j)_{nm}=\phi_{j,n}\phi_{j,m}^*$. Therefore, we obtain in general a lower bound to the rescaled discord as
\begin{equation}\label{Qdq7}
\mathcal{D}_T^{\rm Lo}(\rho)={\max}\left\{0,\,\beta_A\biggl(2-2\sqrt{\frac{\lambda_{1}(\mathbf{A})+\lambda_{2}(\mathbf{A})+\ldots+\lambda_{d_A}(\mathbf{A})}{{\rm Tr}\{\rho^2\}}}\biggl)\right\}.
\end{equation}

%\begin{figure}[b]
%	\begin{center}
%\includegraphics[width=.45\textwidth]{PDTLO.pdf}
%\end{center}
%\caption{Lower bound ${\cal D}_T^{\rm Lo}$ on the rescaled discord [Eq.~(\ref{Qdq7})] plotted versus the true value of the rescaled discord ${\cal D}_T$ [Eq.~(\ref{NormDisc2})] for $10^5$ random two-qubit states. Especially for high values of the rescaled discord, its lower bound is nontrivial and provides an effective estimation of the content of QCs in the considered states. This remains valid  for higher dimensions as well, i.e.~for general $d_A \times d_B$ systems, where a closed computable formula for ${\cal D}_T$ is not available.  \label{figloffa}}
%\end{figure}

%To study the relation between the above lower bound and the actual value of the rescaled discord, we have compared the %two quantities in several examples of bipartite systems. Where a fully analytical solution for the rescaled discord was %not available, we have tackled the problem numerically by generating a large number of random projective measurements, %and choosing the  one that minimizes the distance measure \eqref{TufoDist}. It is found that Eq.~\eqref{Qdq7} is, in many %cases, a strict lower bound to the geometric measure, as shown in Fig.~\ref{figloffa} for a paradigmatic case.

\section{Rescaled discord for two-mode Gaussian states}\label{gaussian}
In this section we study the rescaled discord for the important class of Gaussian states of bipartite CV systems. The Gaussian geometric discord has been introduced in \cite{gauss}, and here, following the same approach, we allow for generalized measurements (POVMs) on the system $A$, restricting them to the class of Gaussian measurements. As the measurements may not be projective it follows that equations \eqref{NormDisc1} and \eqref{NormDisc2} do not hold anymore, thus there is no simple relationship between the Gaussian geometric discord and our rescaled geometric measure.

The restriction to Gaussian measurements allows to obtain analytical closed formulas for both geometric and rescaled discord. Expanding Eq.~\eqref{TufGeoDisc} without the assumption of projective measurements, we obtain
\begin{equation}\label{Gaudi1}
\mathcal{D}_T^{G}(\rho)=\min\left(1-\frac{{\rm Tr}\bigl[\rho\Pi(\rho)\bigl]}{\sqrt{{\rm Tr}[\rho^2]{\rm Tr}\bigl[\Pi(\rho)^2\bigl]}}\right)
=1-\frac{1}{\sqrt{{\rm Tr}\rho^2}}\max\left(\frac{{\rm Tr}\bigl[\rho\Pi(\rho)\bigl]}{\sqrt{{\rm Tr}[\Pi(\rho)^2]}}\right),
\end{equation}
where we have chosen the multiplicative constant $\beta_A=1/2$ for convenience.

Correlations in bipartite Gaussian states depend only on the associated covariance matrix $\Sigma$ and  are invariant under local unitary operations \cite{ourreview}. Focusing on two-mode states, we can simplify the problem and assume covariance matrices written in the standard form
\begin{equation}\label{CovMat}
\Sigma=\left(
                     \begin{array}{cccc}
                       a & 0 & c & 0 \\
                       0 & a & 0 & d \\
                       c & 0 & b & 0 \\
                       0 & d & 0 & b \\
                     \end{array}
                   \right)=\left(
                             \begin{array}{cc}
                               \mathbf{A} & \mathbf{C} \\
                               \mathbf{C} & \mathbf{B} \\
                             \end{array}
                           \right)
\end{equation}
Under a Gaussian POVM on the subsystem $A$, the total covariance
matrix of the bimodal state takes the uncorrelated form
\begin{equation}\label{AftMeasCov}
\sigma=\sigma_A\oplus\sigma_B=\bigl[\mathbf{A}-\mathbf{C}(\mathbf{B}+\sigma_B)^{-1}\mathbf{C}\bigl]\oplus\sigma_B
\end{equation}
where
\begin{equation}\label{CovMeasuB}
\sigma_B=\left(
           \begin{array}{cc}
             m\lambda\cos^2\theta+\frac{m\sin^2\theta}{\lambda} & -\frac{m(\lambda^2-1)\sin\theta\cos\theta}{\lambda} \\
             -\frac{m(\lambda^2-1)\sin\theta\cos\theta}{\lambda} & m\lambda\sin^2\theta+\frac{m\cos^2\theta}{\lambda} \\
           \end{array}
         \right)
\end{equation}
Therefore using the result
${\rm Tr}[\rho_1\rho_2]=1/\sqrt{{\rm Det}[\sigma_1+\sigma_2]/2}$, we arrive at a
compact form for the Gaussian rescaled discord of a bimodal
Gaussian state, which depends only on the elements of the initial
covariance matrix
\begin{equation}\label{Minimizzare}
{\cal D}_T(\rho)=1-({\rm Det}[\Sigma])^{1/4}\max_{\lambda,m,\theta}\biggl(\frac{\sqrt{{\rm Det}[\sigma_A]{\rm Det}[\sigma_B]}}
{{\rm Det}[(\Sigma+\sigma_A\oplus\sigma_B)/2]}\biggl)^{1/2}
\end{equation}
and the maximization is taken over the parameters $0\leq\theta\leq
2\pi$, $\lambda\geq0$ and $m\geq 1$ which characterize the POVM. For
more details we suggest the following references
\cite{AdeDat,GioPa,gauss}.

The maximization can be performed exactly for any physical value of
the parameters $a$, $b$, $c$ and $d$. The general expression is
rather complicated, while a simplified form can be provided in the
case of generally nonsymmetric two-mode squeezed thermal states $\rho_{\rm STS}$, characterized in standard form by $d=\pm c$,
\begin{equation}\label{Asymmtherm}
\mathcal{D}^G_T(\rho_{\rm STS})=1-2\sqrt{\frac{ab-c^2}{2ab+2\sqrt{ab(ab-c^2)}-c^2}}
\end{equation}

\begin{figure}[tb]
	\begin{center}
\includegraphics[width=.4\textwidth]{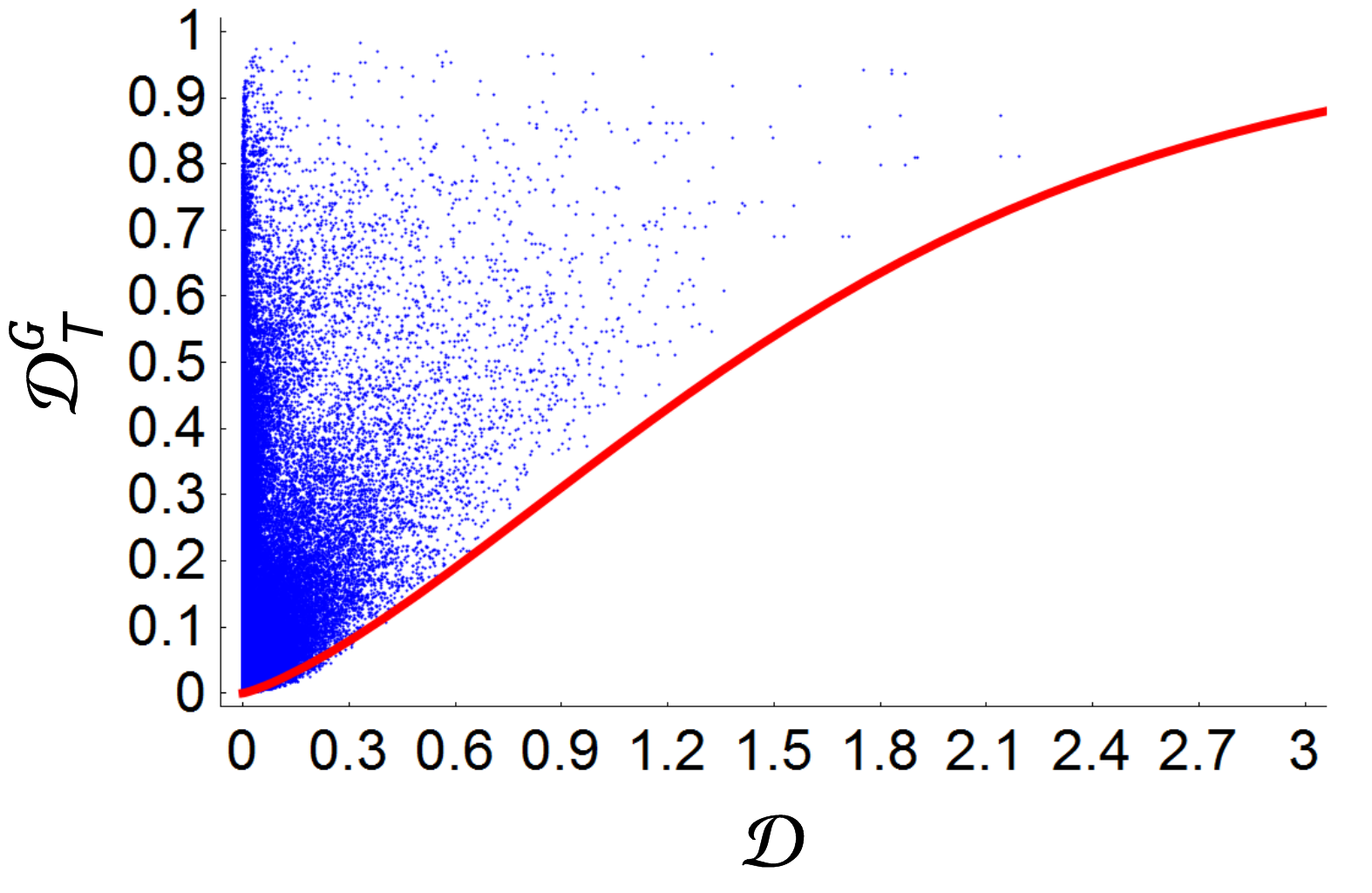}\hspace{.05\textwidth}\includegraphics[width=.4\textwidth]{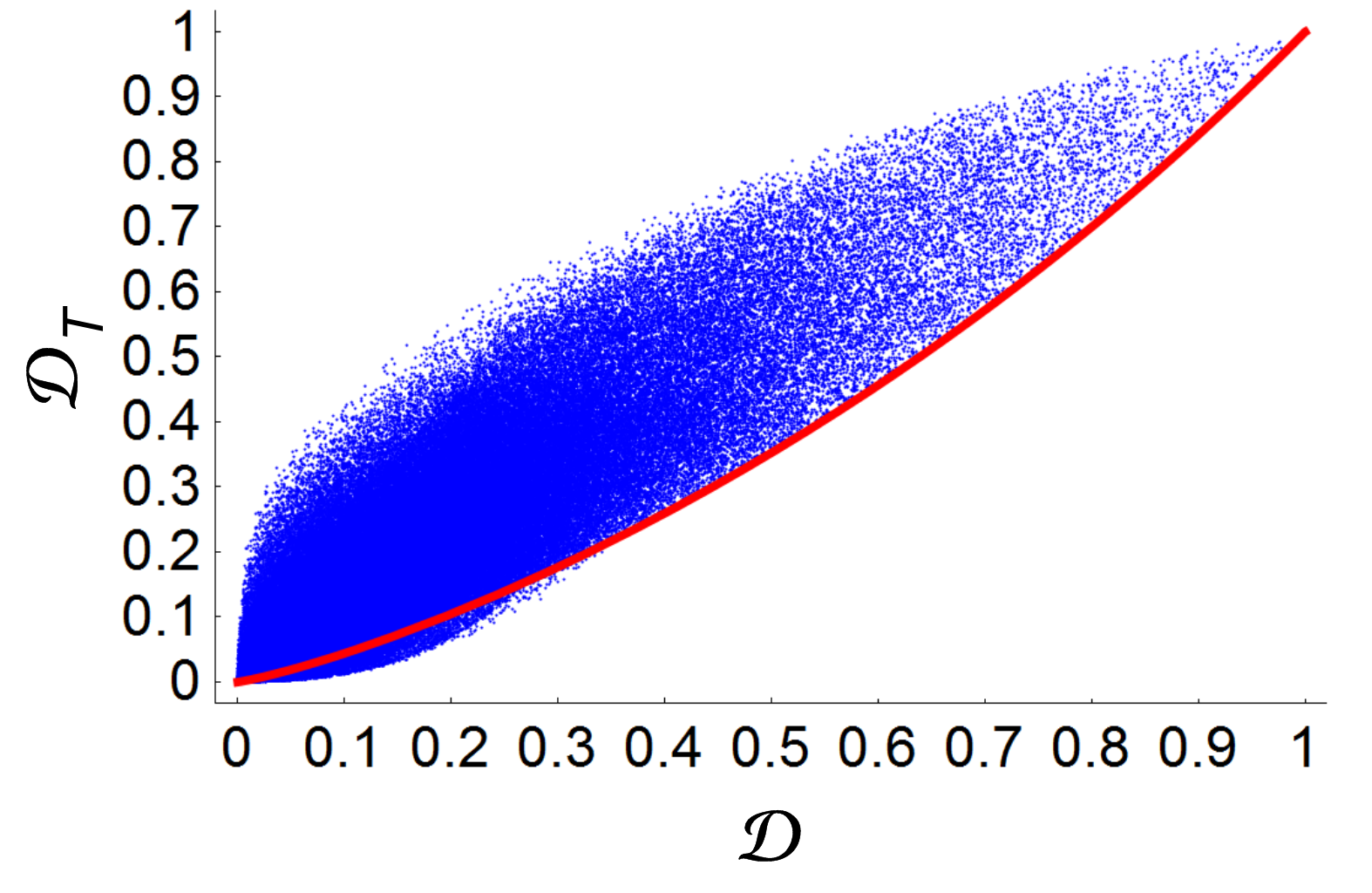}
\end{center}
\caption{(Left) Rescaled discord ${\cal D}_T^G$ restricted to Gaussian measurements versus Gaussian entropic discord ${\cal D}$ \cite{GioPa,AdeDat} for $10^5$ random two-mode Gaussian states; the plot suggests that Gaussian measurements are not optimal to minimize the distance in the definition of the rescaled discord, Eq.~\eqref{TufGeoDisc}, as it can be expected from the analysis of \cite{allegro}. (Right) Rescaled discord ${\cal D}_T$ versus entropic discord ${\cal D}$ for $10^5$ random two-qubit states; the entropic discord is evaluated numerically in this case. In both panels, the red curves accommodate pure states. \label{figpunti}}
\end{figure}

The Gaussian geometric discord defined via the Hilbert-Schmidt metric was known to underestimate the amount of QCs in two-mode Gaussian states compared to the entropic discord \cite{gauss}. It turns out that the Gaussian rescaled discord reverses this behaviour, and in fact overestimates the content of QCs in two-mode Gaussian states. This can be appreciated by comparing ${\cal D}^G_T$ versus ${\cal D}$ for random bimodal Gaussian states, as shown in Fig.~\ref{figpunti}(Left). The origin of this undesired behaviour does not seem to be due to the metric adopted (for example, the use of the Bures distance yields similar results),  but rather to the restriction to Gaussian measurements for the evaluation of ${\cal D}_T$.

As shown in \cite{allegro}, while for the evaluation of the entropic discord it is believed that Gaussian measurements are globally optimal, this is no longer the case for geometric measures of QCs. We expect that ${\cal D}_T$, suitably evaluated by optimizing our distance over general measurements (including, in particular, non-Gaussian measurements such as photon counting), should come down as a reliable estimator for the content of QCs in general two-mode Gaussian states as well, therefore standing in comparison to the entropic ${\cal D}$ similarly to what happens for two-qubit states [Fig.~\ref{figpunti}(Right)]. This will be the subject of further investigation.

\end{document}